\documentclass[twocolumn,secnumarabic,amssymb,nobibnotes,aps,prd,superscriptaddress,nofootinbib,fleqn]{revtex4-2}

\usepackage{amssymb,amsmath,verbatim,mathtools,needspace,enumitem,etoolbox,graphicx,physics,microtype,afterpage,xspace,tabularx,lmodern,multirow,bm,amsthm}
\usepackage{gensymb}
\usepackage[normalem]{ulem}
\usepackage[dvipsnames, usenames]{xcolor}
\usepackage{xr-hyper}
\definecolor{linkcolor}{rgb}{0.0,0.3,0.5}
\usepackage[unicode, colorlinks=true, linkcolor=linkcolor, citecolor=linkcolor, filecolor=linkcolor, urlcolor=linkcolor, linktocpage, breaklinks]{hyperref}
\usepackage[all]{hypcap}
\usepackage[T1]{fontenc}
\usepackage[utf8]{inputenc}
\usepackage[usenames,dvipsnames]{xcolor}
\hypersetup{colorlinks=true,citecolor=romared,linkcolor=romared,urlcolor=romared}

\definecolor{romared}{RGB}{142,0,28}
\newcommand{\nn}{\nonumber}

\newcommand{\be}{\begin{equation}}
\newcommand{\ee}{\end{equation}}

\def\be{\begin{equation}}
\def\ee{\end{equation}}
\newcommand{\beq}{\begin{eqnarray}}
\newcommand{\eeq}{\end{eqnarray}}

\usepackage{makecell}
\usepackage{soul}
\newcommand{\iu}{\mathrm{i}}
\newcommand{\inn}{\mathrm{in}}
\newcommand{\out}{\mathrm{out}}
\newcommand{\sign}{\mathrm{sign}}

\makeatletter
\newcommand*{\addFileDependency}[1]{%
  \typeout{(#1)}
  \@addtofilelist{#1}
  \IfFileExists{#1}{}{\typeout{No file #1.}}
}
\makeatother

\newcommand*{\myexternaldocument}[1]{%
    \externaldocument{#1}%
    \addFileDependency{#1.tex}%
    \addFileDependency{#1.aux}%
}

\myexternaldocument{Supp}

\begin{document}

\title{On energy extraction from Q-balls and other fundamental solitons}
\author{Vitor Cardoso} 
\affiliation{Niels Bohr International Academy, Niels Bohr Institute, Blegdamsvej 17, 2100 Copenhagen, Denmark}
\affiliation{CENTRA, Departamento de F\'{\i}sica, Instituto Superior T\'ecnico -- IST, Universidade de Lisboa -- UL, Avenida Rovisco Pais 1, 1049-001 Lisboa, Portugal}
\author{Rodrigo Vicente} 
\affiliation{Institut de Fisica d'Altes Energies (IFAE), The Barcelona Institute of Science and Technology,
Campus UAB, 08193 Bellaterra (Barcelona), Spain}
\author{Zhen Zhong} 
\affiliation{CENTRA, Departamento de F\'{\i}sica, Instituto Superior T\'ecnico -- IST, Universidade de Lisboa -- UL, Avenida Rovisco Pais 1, 1049-001 Lisboa, Portugal}

\date{\today}

\begin{abstract}
Energy exchange mechanisms have important applications in particle physics, gravity, fluid mechanics, and practically every field in physics.
In this letter we show, both in frequency and time domain, that energy enhancement is possible for waves scattering off fundamental solitons (time-periodic localized structures of bosonic fields), without the need for rotation nor translational motion.
We use two-dimensional Q-balls as a testbed, providing the correct criteria for energy amplification, as well as the respective amplification factors, and we discuss possible instability mechanisms. Our results lend support to the qualitative picture drawn in Ref.~\cite{Saffin:2022tub}; however we show that this enhancement mechanism is not of superradiant-type, but instead a ''blueshift-like`` energy exchange between scattering states induced by the background Q-ball, which should occur generically for any time-periodic fundamental soliton. This mechanism does not seem to lead to instabilities. 
\end{abstract}

\maketitle

\noindent{\bf \em Introduction.} 
Energy exchange phenomena play a pivotal role in the small and large scale dynamics of all observed phenomena. A class of these are termed ``superradiant'' and take place when an object with many internal degrees of freedom -- which can internally dissipate energy -- is able to amplify certain impinging radiation modes, while increasing its internal energy~\cite{Bekenstein:1998nt}. In this mechanism the necessary energy to enhance the radiation (and increase the object's internal energy) is usually provided by kinetic energy. For objects in uniform linear motion superradiance requires a velocity larger than the characteristic phase velocity in the medium (e.g., Vavilov-Cherenkov effect), while rotational superradiance requires angular velocities larger than the characteristic (angular) phase velocity in the medium~\cite{zeldovich1,zeldovich2,Bekenstein:1998nt,Brito:2015oca}. A different energy exchange mechanism is realised by the interaction of radiation with a time-dependent background. Time-periodic backgrounds tend to induce a coupling between a discrete set of modes, leading to an effective energy exchange with radiation. In particular, the energy of radiation may be enhanced through a blue-shift of incoming modes to higher frequencies (as it happens with oscillating cavity walls and objects~\cite{Dittrich:1994nk,Cooper:1993,PhysRevLett.73.1931,Ikeda:2021pkb} or moving objects~\cite{Cardoso:2019dte}).

Superradiance has attracted a considerable amount of attention in black hole physics, since it may be a viable way to power violent phenomena in the cosmos, or even to transfer energy between black holes and new fundamental degrees of freedom~\cite{Brito:2015oca}. Spinning black holes have two properties which are ideal for superradiance: an ergoregion that effectively couples radiation to spacetime, and an horizon that quenches negative-energy modes, allowing for energy exchange in a stable manner~\cite{Vicente:2018mxl}. It was recently claimed that Q-balls -- a type of non-topological scalar field soliton~\cite{Coleman:1985ki,Tsumagari:2009zp} -- are also prone to superradiance, not requiring rotation nor any type of motion in real space~\cite{Saffin:2022tub}. 

Here, we argue that even though the original proof was flawed, energy extraction from Q-balls is indeed possible. It is not of superradiant nature, but it involves rather a blue-shift mechanism powered by the time-periodic background, akin to Doppler shift of radiation in oscillating cavities. 
Due to Derrick's theorem~\cite{Derrick:1964ww} -- showing that there is no stable time-independent solution of finite energy for a wide class of nonlinear wave equations -- most fundamental solitons are expected to be time-periodic. This energy extraction mechanism most likely extends to all these objects, since nonlinearites will induce the necessary mode-mixing in impinging radiation. We have explicitly proved that this is the case for (Newtonian) boson stars, which have attracted attention in connection with dark matter physics and black hole mimickers~\cite{Liebling:2012fv,Cardoso:2019rvt}.\footnote{We use geometric units with~$G=c=1$ and a mostly positive signature for the spacetime metric throughout.}

\noindent{\bf \em Q-balls as a testbed.} 
Consider a simple U(1)-symmetric theory of a complex scalar field~$\bar{\Phi}$ in a three-dimensional flat spacetime~$(\mathbb{R}^{3},\eta_{\alpha \beta})$ described by the action
\begin{equation}\label{eq:action}
S=-\tfrac{1}{2}\int \mathrm{d}^{3}\bar{x} \sqrt{-\eta}\left[\partial^{\bar{\alpha}} \bar{\Phi}^*\partial_{\bar{\alpha}} \bar{\Phi}+V\left(|\bar{\Phi}| \right) \right] ,
\end{equation}
where~$\eta \coloneqq \mathrm{det} (\eta_{\alpha \beta})$, with the potential~$V=\mu^2 |\bar{\Phi}|^2-\lambda |\bar{\Phi}|^4+\bar{g} |\bar{\Phi}|^6$. We work with the re-scaled the dimensionless quantities~$x\coloneqq \mu \bar{x}$, $\Phi \coloneqq \sqrt{\lambda} \bar{\Phi}/\mu$ and~$g\coloneqq \mu^2 \bar{g}/\lambda^2$, in terms of which the potential reads~$V=|\Phi|^2-|\Phi|^4+g |\Phi|^6$. We restrict to~$g\geq 1/4$ (so that~$\Phi=0$ is the true vacuum).
The field satisfies the equation of motion
\begin{equation}
    \Box_\eta \Phi - \frac{\partial V}{\partial \Phi^*}=0,\label{eq:eom}
\end{equation}
and possess the divergenceless Noether current
\begin{equation}
J_{Q}^\alpha=\Im\left(\Phi^* \partial^\alpha \Phi \right), 
\end{equation}
and energy current (as measured by a family of parallel inertial observers with 4-velocity~$\delta_t^\alpha$)
\begin{equation}
J_E^\alpha=-\eta^{\alpha \beta}T_{\beta \gamma} \delta_t^\gamma, 
\end{equation}
with the (also divergenceless) energy-momentum tensor
\begin{equation}
T_{\alpha \beta}=\partial_{(\alpha}\Phi^* \partial_{\beta)}\Phi-\tfrac{1}{2}\eta_{\alpha \beta}\left[\partial^\gamma \Phi^* \partial_\gamma \Phi + V\left(|\Phi|\right) \right].
\end{equation}

\begin{figure}[]\label{fig:qball}
\includegraphics[width=1.1\linewidth]{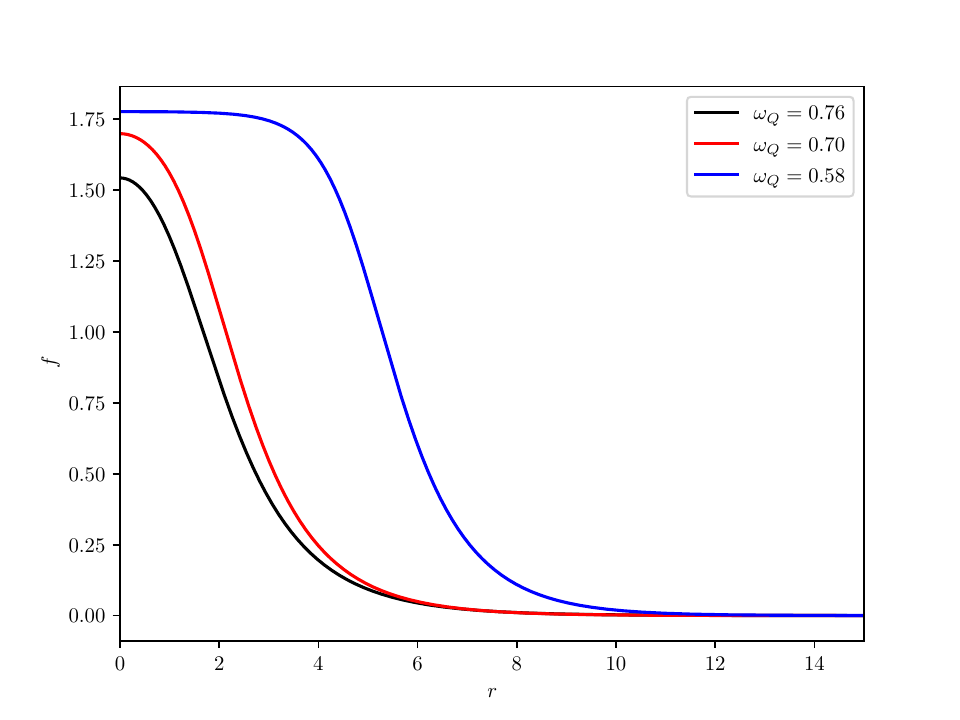}
\caption{Radial profile~$f(r)$ of the Q-balls studied in this work, with~$\omega_Q = (0.58,0.70,0.76)$,~$m_Q=0$ and~$g=1/3$. The values at the origin are, respectively,~$f(0)=(1.78,1.70,1.54)$, their charge is~$Q=\int_\Sigma J^t_Q =(70.77,13.70,9.76)$ and energy is~$E=\int_\Sigma J_E^t=(47.02,12.00,9.14)$. We consider only ground-state (nodeless) Q-ball solutions.}
\end{figure}
Using polar coordinates~$(r,\varphi)$ on the hypersurfaces~$\Sigma_t$ orthogonal to~$\delta_t^\alpha$, Q-balls are solutions of the form
\begin{equation}
\Phi_Q(t,r,\varphi)=\frac{1}{\sqrt{2}} f(r) e^{-\iu \left( \omega_Q t- m_Q \varphi\right)},
\end{equation}
that are regular at~$r=0$ and~$r=+\infty$ (without loss of generality, we consider~$\omega_Q>0$ and~$m_Q\in \mathbb{Z}_0^+$). Some radial profiles with different~$\omega_Q$ are shown in Fig.~\ref{fig:qball}. To exist the Q-ball frequency must be~$\omega_Q^{\textrm{thin}}<\omega_Q<1$, with the lower bound~$\omega_Q^{\textrm{thin}}\coloneqq \textrm{min}[2 V/|\Phi|^2]$ corresponding to the so-called \emph{thin-wall} limit~\cite{Coleman:1985ki}.

\noindent{\bf \em Linear perturbations.} 
A sufficiently small perturbation~$\Phi_1$ to a Q-ball background solution, $\Phi \approx \Phi_Q+\Phi_1$ with~$|\Phi_1|\ll |\Phi_Q|$, satisfies the \emph{linearized} equation of motion
\begin{equation}
\Box_\eta \Phi_1 - U(r) \Phi_1-e^{-2 \iu \left(\omega_Q t - m_Q \varphi\right)}W(r)\Phi_1^*=0, \label{eq:eom_perturb}
\end{equation}
where~$U\coloneqq  1-2 f^2+\frac{9}{4} g f^4$ and~$W\coloneqq - f^2+\frac{3}{2} g  f^4$. The last term gives rise to mode-mixing and, thus, this equation does not admit monochromatic solutions. 

\noindent{\bf \em Frequency-domain analysis.} 
The solutions with minimal frequency-content are of the form
\begin{equation}
\Phi_1=\phi_+(r) e^{- \iu \left(\omega_+ t- m_+ \varphi \right)}+\phi_-(r) e^{- \iu \left(\omega_- t- m_- \varphi \right)},\label{eq:Phi_1}
\end{equation}
where~$\omega_{\pm}= \omega_Q \pm \omega$ and~$m_\pm = m_Q \pm m$. The mode functions~$(\phi_+,\phi_-)$ satisfy the \emph{coupled} system
\begin{equation}\label{mode_funct}
\frac{1}{r}\partial_r\left(r\partial_r \phi_\pm\right)+\left[\omega_\pm^2-U-\frac{m_\pm^2}{r^2}\right]\phi_\pm-W \phi_\mp^* =0,
\end{equation}
which clearly shows the mode-mixing between~$\phi_+$ and~$\phi_-$ introduced by the background Q-ball. 
At infinity the mode functions are
\begin{equation} \label{phi_inf}
\lim_{r\to \infty} \phi_\pm \approx \tfrac{1}{\sqrt{2\pi|k_\pm| r}}\left(A_{\pm}^{\out} e^{\iu k_\pm r}+ A_\pm^{\inn} e^{-\iu k_\pm r}\right),
\end{equation}
and at the origin
\begin{equation}\label{regular_0}
\lim_{r\to 0}\phi_\pm \approx C_\pm \left(|k_\pm| r \right)^{|m_\pm|},
\end{equation}
where~$k_\pm\coloneqq \pm s_{\omega}(\omega_\pm^2-1)^{1/2}$ with~$s_\omega \coloneqq \sign(\omega)$.
From the scale invariance of the linear perturbations, we can choose $C_+=1$ without loss of generality.

In this work we focus on perturbations satisfying the double condition~$|\omega_\pm|>1$, in which case both modes describe propagating waves (i.e., \emph{scattering states}); the waves with amplitude~$A_\pm^\out$ (resp.,~$A_\pm^\inn$) have radial group velocity~$\frac{\mathrm{d} \omega_\pm}{\mathrm{d}k_\pm}$ (resp.,~$- \frac{\mathrm{d} \omega_\pm}{\mathrm{d}k_\pm}$). This means that~$A_\pm^\out$ describe outgoing states propagating in the~$\delta_r^\mu$ direction, the opposite to the direction of propagation of incoming~$A_\pm^\inn$ states.

The time-averaged flux of $Q$-charge through a 1-sphere of radius~$r$ (with~$r\to \infty$) is
\begin{align}
\mathcal{F}_Q&= \lim_{r\to \infty} r \int_0^{2\pi} \mathrm{d}\varphi \left< J^r_Q\right> \nn \\
&\approx  s_\omega \sum_{s=+,-} s\left(|A_s^\out|^2-|A_s^\inn|^2\right),
\end{align}
that of of energy is
\begin{align}
    \mathcal{F}_E&= \lim_{r\to \infty} r \int_0^{2\pi} \mathrm{d}\varphi \left< J^r_E\right> \nn\\
    &\approx s_\omega \sum_{s=+,-}s\,\omega_s \left(|A_s^\out|^2-|A_s^\inn|^2\right),
\end{align}
and of angular momentum is
\begin{align}
    \mathcal{F}_L&= \lim_{r\to \infty} r \int_0^{2\pi} \mathrm{d}\varphi \left< T_{r \varphi}\right> \nn\\
    &\approx s_\omega \sum_{s=+,-}s\,m_s \left(|A_s^\out|^2-|A_s^\inn|^2\right).
\end{align}
where~$\left<\,\cdot \,\right>\equiv \lim_{T\to \infty} \frac{1}{T}\int_0^T \mathrm{d}t\, (\,\cdot\,)$.

It is easy to see from Eq.~\eqref{mode_funct} that the quantity
\begin{equation}
    J_\phi\coloneqq r \Im\left(\phi_+^*\partial_r \phi_+ +\phi_- \partial_r\phi_-^{*}\right)\,,
\end{equation}
is independent of~$r$ on-shell, i.e.,~$\partial_r J_\phi=0$ for a solution of Eq.~\eqref{mode_funct}. Regularity of the linear perturbations at the origin [c.f. Eq.~\eqref{regular_0}] implies then that~$J_\phi=0$. But, since~$J_\phi(r\to \infty)\propto\sum_{s=\pm}(|A_s^\out|^2-|A_s^\inn|^2)$, one finds
\begin{equation}\label{eq:q_conserv}
    |A_-^\out|^2+|A_+^\out|^2=|A_-^\inn|^2+|A_+^\inn|^2.
\end{equation}
The time-averaged fluxes can then be written as
\begin{gather}
    \mathcal{F}_Q \approx 2 s_\omega \left(|A_+^\out|^2-|A_+^\inn|^2\right), \\
    \mathcal{F}_E \approx 2 s_\omega \omega_Q \left(|A_+^\out|^2-|A_+^\inn|^2\right),\\
    \mathcal{F}_L \approx 2 s_\omega m_Q \left(|A_+^\out|^2-|A_+^\inn|^2\right).
\end{gather}
Thus, both~$Q$-charge, energy and angular momentum can be exchanged with the background (exchange of angular momentum is only possible with a spinning Q-ball).
Energy extraction from a spinning Q-ball is necessarily accompanied by angular momentum extraction.

\begin{figure}[h]\label{fig:Z_E}
  \centering
  \includegraphics[width=1.1\linewidth]{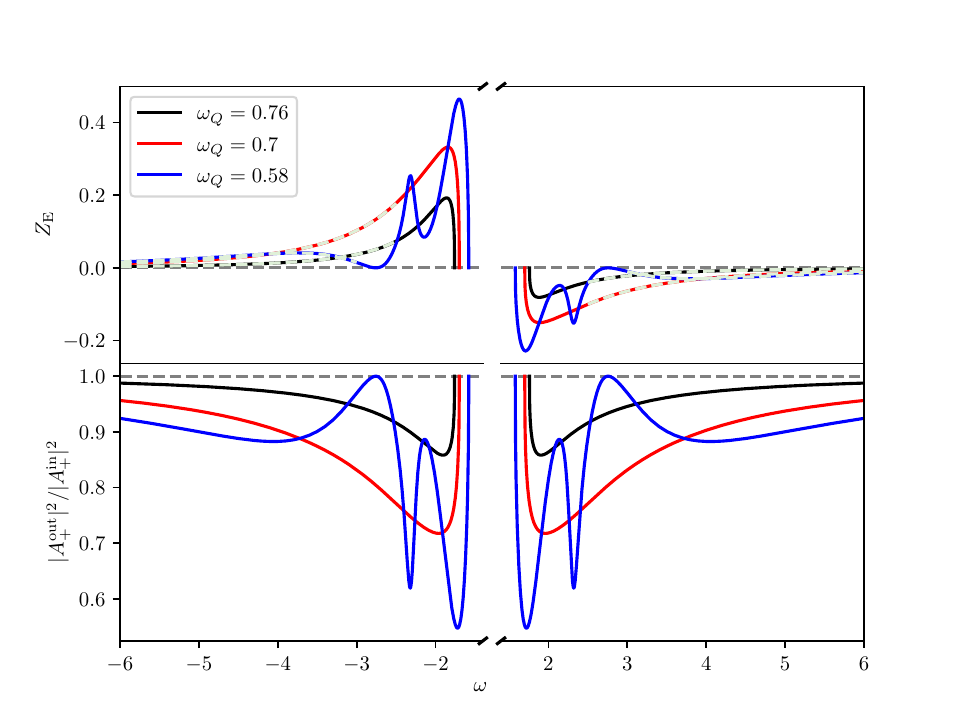}
  \caption{\emph{Upper panel:} Relative energy amplification factor, $Z_E$, of an incoming mode~$\phi_+$ with~$m=0$ scattering off a non-spinning Q-ball; the coupling is~$g = 1/3$. The dashed lines are the results from time-domain simulations. 
  \emph{Lower panel:} Ratio~$|A_+^\out|^2/|A_+^\inn|^2$, which is seen to be~$\leq1$ for all~$\omega$. %
  } 
\end{figure}
The amplification factors of~$Q$-charge, energy and angular momentum in the scattering process, defined as (the absolute value of) the ratio of the time-averaged outgoing flux to the incoming flux, are respectively
\begin{gather}
    1+Z_Q=\left|\frac{|A_+^\out|^2-|A_-^\out|^2}{|A_+^\inn|^2-|A_-^\inn|^2}\right|,\\
    1+Z_E=\left|\frac{\omega_+ |A_+^\out|^2-\omega_- |A_-^\out|^2}{\omega_+ |A_+^\inn|^2-\omega_-|A_-^\inn|^2}\right|,\label{eq:Z_E}\\
    1+Z_L=\left|\frac{m_+ |A_+^\out|^2-m_-|A_-^\out|^2}{m_+ |A_+^\inn|^2-m_- |A_-^\inn|^2}\right|\label{eq:Z_L}.
\end{gather}
The $Z$-factors measure the \emph{relative} amplification (or attenuation) in the scattering process: $Z>0$ for amplification, whereas $Z<0$ for attenuation.

Note that asymptotic fluxes are the appropriate quantities that allow one to discriminate between amplification and attenuation in a scattering process. In Ref.~\cite{Saffin:2022tub} an alternative criteria for energy amplification was used, based instead on (the absolute value of) the ratio of energy contained in outgoing states to incoming ones in some asymptotic annular region~$r_1<r<r_2$. Given the different propagation speeds of the~$\phi_\pm$ modes, this is not an appropriate measure. In fact, in the limit where one mode has arbitrarily small group velocity, its energy density becomes arbitrarily large (c.f. Fig.~2 of \cite{Saffin:2022tub}), but its energy flux -- which determines the rate of energy exchanged with the exterior -- may still be small.

From Eq.~\eqref{eq:Z_E} it is clear that energy amplification will occur for any incoming state with~$A_+^\inn=0$ and~$\omega>0$, or~$A_-^\inn=0$ and~$\omega<0$. On the other hand, there will be energy attenuation ($Z_E<0$) for any incoming state with~$A_-^{\inn}=0$ and~$\omega>0$, or~$A_+^{\inn}=0$ and~$\omega<0$. These sufficient conditions for energy amplification do \emph{not} agree with the ones found in Ref.~\cite{Saffin:2022tub}, due to the different criteria for amplification given there. The outcome of a more general scattering process in which the incoming state contains a mixture of~$\phi_+$ and~$\phi_-$ modes depends on the details of the process, i.e., on the scattering parameters~$\omega$,~$m$ and~$A_-^\inn/A_+^\inn$. 
Similarly, one can immediately see from Eq.~\eqref{eq:Z_L} that for a spinning Q-ball ($m_Q>0$) there is angular momentum amplification for an incoming state with~$A_+^\inn=0$ and~$m\geq m_Q$, or~$A_-^\inn=0$ and~$m\leq - m_Q$, and attenuation for an incoming state with~$A_-^\inn=0$ and~$m\geq m_Q$, or~$A_+^\inn=0$ and~$m\leq - m_Q$. 

The energy extraction mechanism discussed here is \emph{not} of superradiant-type. In fact, enhancement of energy (or~$Q$-charge, or angular momentum) of a single incoming state is never observed, as can be seen by noting that~$|A_s^\out|/|A_s^\inn|\leq 1$, either directly from Eq.~\eqref{eq:q_conserv}, or from the lower panel of Fig.~\ref{fig:Z_E}.
Instead, the energy extraction is accomplished through a ''blueshift-like`` exchange, where the time-dependent background effectively pumps energy from the lowest energy (i.e., frequency) state to the highest. As we discuss in the next section, the Q-ball will then evolve to a new Q-ball with different parameters as to conserve the total~$Q$-charge, energy, and angular momentum in the process.

The~$Z_E$-factor is shown in the upper panel of Fig.~\ref{fig:Z_E} for an incoming~$\phi_+$ mode (i.e.,~$A_-^\inn=0$) from solving numerically Eq.~\eqref{mode_funct}; these results show a remarkable agreement with those from evolving a wavepacket in the time-domain (discussed in the next section). As expected, we find energy amplification of~$\omega<0$ modes for all Q-balls, and attenuation of~$\omega>0$ modes. Note also how energy extraction is more effective for Q-balls closer to the thin-wall limit, for which the field profile is nearly constant in the interior (c.f. Fig.~\ref{fig:qball}).

\begin{figure}[t]
\includegraphics[width=1.1\linewidth]{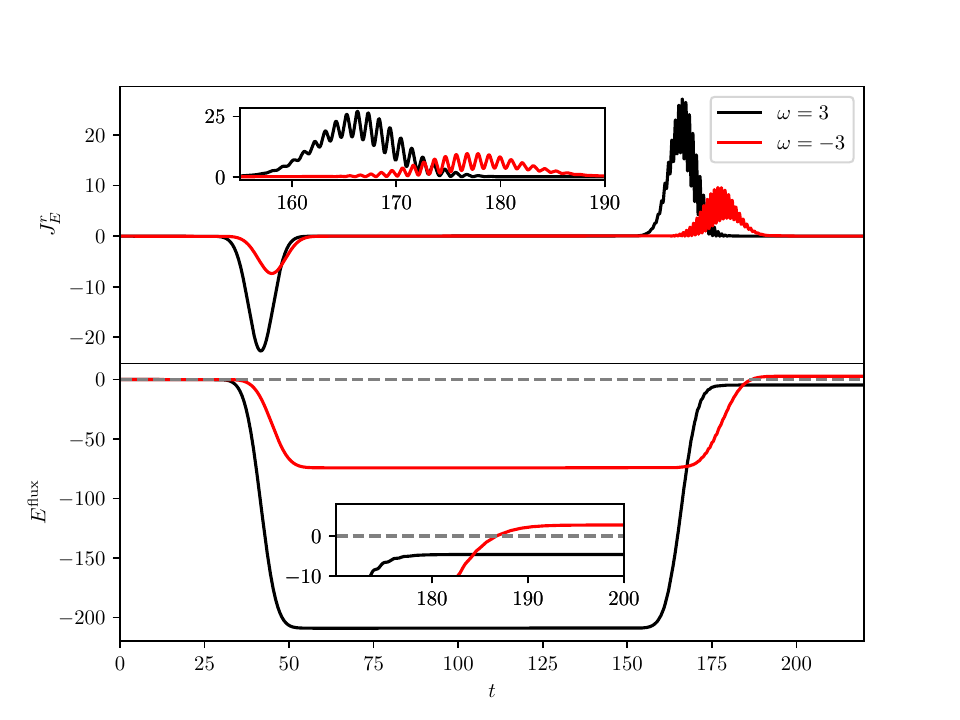}
\caption{%
Wavepacket scattering off a non-spinning Q-ball with~$\omega_Q = 0.76$; the coupling is~$g = 1/3$ and we use initial conditions with~$\sigma_r = 5$,~$r_0 = 100$, and $\omega_0 =\{-2.24,3.76\}$.
\emph{Upper panel:} Energy flux $J^r_E(r=60)$ as function of time, with the outgoing flux zoomed-in. \emph{Lower panel:} Integrated energy flux $E^{\mathrm{flux}} = \int_0^t \mathrm{d}t J^r_E(r=60)$, with an inset showing attenuation ($Z_E = -0.022$) for $\omega = 3$ and amplification ($Z_E = 0.037$) for $\omega = -3$. These results are in excellent agreement with the frequency-domain analysis (c.f. Fig.~\ref{fig:Z_E}). 
}
\label{fig:timedomain}
\end{figure}

\noindent{\bf \em Time-domain analysis.}
To verify the consistency of our results, we performed a time-domain evolution of a~$m=0$ wavepacket scattering off a Q-ball, solving Eq.~\eqref{eq:eom_perturb} with initial conditions appropriate for a Gaussian wavepacket with (average) frequency~$\omega_0$ and radial width~$\sigma_r$,
\begin{gather}
\Phi_1(0, r) = e^{- \frac{\left(r - r_0\right)^2}{2 \sigma_r^2} } e^{-   \iu\, s_{\omega_0}\sqrt{\omega_0^2-1}\, r}\,,\\
\partial_t \Phi_1(0, r) = -\iu\, \omega_0 \Phi_1(0, r).
\end{gather}
Additionally, we have also evolved the full \emph{nonlinear} Eq.~\eqref{eq:eom} around non-spinning Q-ball solutions, using the initial conditions
\begin{gather}
\Phi(0, r) = \Phi_Q(0,r)+\delta\, e^{- \frac{\left(r - r_0\right)^2}{2 \sigma_r^2} } e^{-   \iu s_{\omega_0}\sqrt{\omega_0^2-1}\, r}\,,\\
\partial_t \Phi(0, r) =-\iu\left[ \omega_Q \Phi_Q + \omega_0 (\Phi-\Phi_Q)\right](0, r).
\end{gather}
For small enough~$\delta\ll \Phi_Q(0,0)$ the nonlinear scattering results are consistent with the linearized calculations.
We employ a fourth-order finite difference scheme in space and a standard Runge-Kutta 4 algorithm in time.
The results are summarized in Figs.~\ref{fig:Z_E}--\ref{fig:timedomain}.
Figure~\ref{fig:timedomain} shows how a smooth nearly monochromatic wavepacket scatters off and acquires extra energy. For an incoming wavepacket with~$\omega_0= 3.76$, a Fourier transform shows that the scattered wavepacket contains the modes~$\omega_- \approx -2.24$ and~$\omega_+ \approx 3.76$, in agreement with the analysis around Eq.~\eqref{mode_funct} for~$\omega=3$. The calculation of the amplification factor~$Z_E$ from time-domain data is straightforward, and the results are shown in Fig.~\ref{fig:Z_E} together with the frequency-domain prediction. The agreement is remarkable.

The fact that the currents~$J_Q^\alpha$ and~$J_E^\alpha$ are divergenceless, implies the conservation of both~$Q$-charge and energy; so, any change in~$Q$ and~$E$ of the scattering states, must be accompanied by a symmetric change in the Q-ball. Indeed, our results from full nonlinear evolutions show that, after the scattering takes place, the remnant relaxes to a configuration compatible with a \emph{new} Q-ball with parameters such that charge and energy conservation are verified. This allows us to predict the maximum extractable energy from these Q-balls. We verified numerically that the entire family of Q-balls in our testbed model (with~$g=1/3$) is \emph{absolutely stable} (i.e., have~$E/Q<1$), and their energy decreases monotonically to~$E\approx 5.85$ in the limit~$\omega \to 1$; this is a qualitative agreement with the analytic approximation of Ref.~\cite{Multamaki:1999an}. Thus, the maximum extractable energy for the Q-balls with~$\omega_Q=(0.58,0.70,0.76)$ is, respectively,~$(88\%,51\%,36\%)$ of their initial energy. One can then -- in principle -- extract more energy from these objects than from spinning BHs~\cite{Christodoulou:1971pcn}.

\noindent{\bf \em Other fundamental objects.}
The simplicity of the energy extraction mechanism discussed in this work indicates that other fundamental objects (like boson stars~\cite{Liebling:2012fv}, Proca stars~\cite{Brito:2015pxa}, etc) might be prone to the same mechanism.
For instance, the extension of the proof to spherically symmetric Newtonian boson stars is particularly simple. It goes as follows. 
Consider a four-dimensional complex scalar field~$\Phi$ of mass~$\mu$ minimally coupled to gravity. In the non-relativistic limit the scalar field~$\Phi\coloneqq e^{-\iu \mu t} \Psi/\sqrt{\mu}$ satisfies the Schrödinger-Poisson system~\cite{Annulli:2020lyc}
\begin{gather}
    \iu \partial_t\Psi=-\tfrac{1}{2\mu} \nabla_i\nabla^i\Psi+\mu U \Psi, \label{eq:KG} \\
    \nabla_i\nabla^i U=4 \pi \mu |\Psi|^2.
\end{gather}
Newtonian boson stars are solutions of the form~$\Psi_0=\psi_0(r) e^{-\iu \omega_0 t}$ with~$0<\omega_0<\mu$, and a static gravitational potential~$u_0(r)$, both satisfying regularity at~$r=0$ and~$r=+\infty$. Inspired by the Q-ball treatment, it is not hard to see that:
\begin{enumerate}
    \item The linearized system contains time-periodic coefficients, and all of these are of the form~$\propto e^{-2\iu \omega_0 t}\Psi_1^*$ and~$\propto e^{-\iu \omega_0 t}U_1$.
    \item This leads to mode-mixing between the states~$\omega_-=\omega_0-\omega_1$ and~$\omega_+=\omega_0+\omega_1$; so, the minimal frequency-content solutions~$\Psi_1$ must contain these two modes [c.f. Eq.~\eqref{eq:Phi_1}], while~$U_1$ is real-valued and has frequencies~$\pm \omega_1$. This was also seen, e.g., in Ref.~\cite{Macedo:2013jja,Annulli:2020lyc}.
    \item The current 
    \begin{align}
        J_{\psi,u}\coloneqq r^2 \Im&\big(\psi_+^*\partial_r \psi_+ +\psi_- \partial_r\psi_-^{*}\nn\\
        &-\tfrac{\mu}{2\pi}u_+ \partial_r u_+ -\tfrac{\mu}{2\pi}u_- \partial_r u_-\big)
    \end{align}
    satisfies~$\partial_r J_{\psi,u}=0$. Regularity at the origin implies~$J_{\psi,u}=0$, and the asymptotic behaviour gives~$J_{\psi,u}(r\to \infty)\propto\sum_{s=\pm}(|A_s^\out|^2-|A_s^\inn|^2)$.
    \item As for the Q-ball, an incoming state with~$A_+^\inn=0$ and~$\omega_1>0$, or~$A_-^\inn=0$ and~$\omega_1<0$ will necessarily extract energy from the boson star.
\end{enumerate}

\noindent{\bf \em Discussion.}
Our results establish that energy extraction from Q-balls is allowed by a process that does not require rotation nor any motion in real space, but merely an interaction with a time-periodic background. As we showed, there is no superradiance in this process, since an incoming single state is never enhanced. Instead, this mechanism bears strong similarities to the blue-shift mechanism reported when a wave is trapped within a cavity whose boundaries oscillate periodically~\cite{Dittrich:1994nk,Cooper:1993,PhysRevLett.73.1931,Ikeda:2021pkb}. In general, an interaction of radiation with a time-periodic background tends to lead to mode-mixing, effectively coupling a discrete set of modes. If initially the radiation has only support in the lowest frequency (i.e., energy) mode, the interaction with the background will then re-distribute radiation over the several coupled modes, leading to an overall energy amplification of radiation powered by the internal energy of the background. The simplicity of this argument indicates that other types of fundamental solitons -- expected to be time-dependent due to Derrick's theorem -- might be prone to the same mechanism. We have showed it explicitly for Newtonian boson stars.

If a continuous family of (stable) soliton solutions exists (as for Q-balls, boson and Proca stars) the backreaction of the scattering process on the soliton is straightforward. For instance, if we continuously ''illuminate`` a Q-ball with a single state~$\phi_+$ of frequency~$\omega_+>1$, the deposition of energy in the object will drive secularly the soliton through a continuous sequence of Q-balls with increasingly larger energy and charge (i.e., along the thin-wall limit). The process will eventually saturate if~$\omega_-=-1$ is attained. Similarly, illuminating a Q-ball with a single state~$\phi_+$ of frequency~$\omega_+<-1$ will continuously extract energy and charge from the object driving the soliton through a continuous sequence of Q-balls with increasingly smaller energy and charge, which saturates if~$\omega_+=-1$ is attained. The evolution picture drawn here is compatible with our full nonlinear time evolutions of wavepacket scattering. Note however that this evolution may be hindered by the soliton reaching some unstable configuration.

\begin{figure}[t!]
  \centering
  \includegraphics[width=1.1\linewidth]{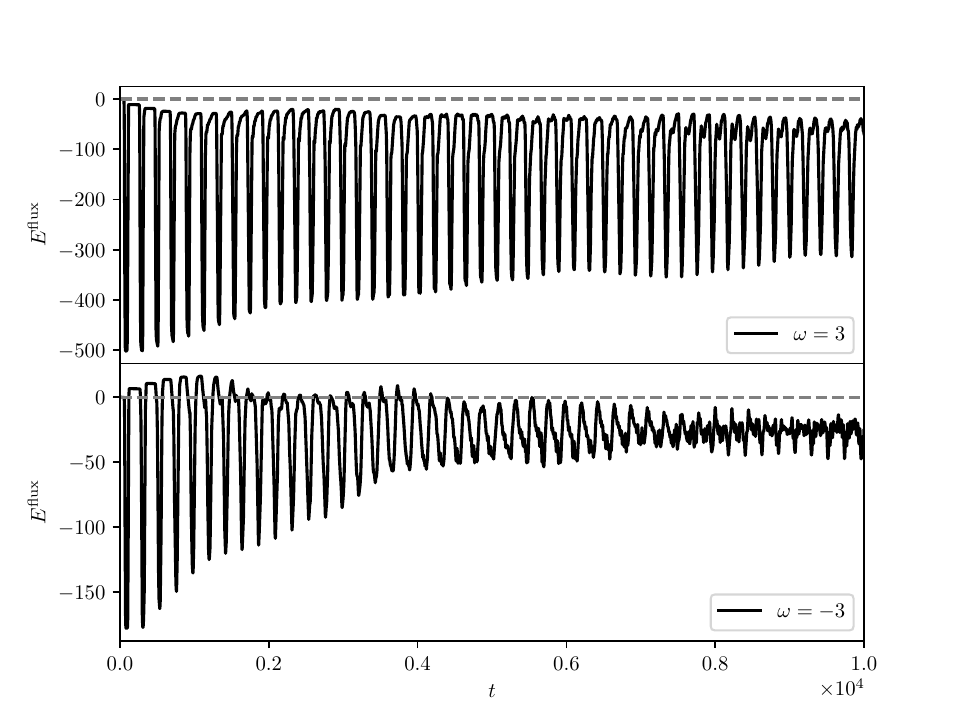}
  \caption{Integrated energy flux~$E^{\mathrm{flux}} = \int_0^t \mathrm{d}t J^r_E$ at~$r = 20$ for a wavepacket scattering off a non-spinning Q-ball with~$\omega_Q = 0.76$ inside a cavity; the coupling is~$g= 1/3$ and we use initial conditions with~$\sigma_r = 5$,~$r_0 = 80$, and~$\omega_0 = \{-2.24,3.76\}$. We impose Dirichlet conditions~$\Phi_1=0$ at~$r = 100$ to model the cavity.\label{fig:cavity}
  }
\end{figure}

In the case of superradiance, energy extraction can also be associated to instabilities when the radiation is confined and forced to repeatedly interact with the amplifying body (the cavity can be artificial, or arse naturally as part of the geometry)~\cite{Dittrich:1994nk,Cooper:1993,PhysRevLett.73.1931,Ikeda:2021pkb,Cardoso:2004nk,Cardoso:2004hs,Brito:2015oca}. However, for the energy extraction mechanism discussed in this work, it is not as obvious (as it is for superradiance) that confining the radiation should lead to an instability. This is because, even if the initial (incoming) state is chosen so that its energy must be necessarily amplified, the scattered state will be in a mixture of modes whose energy does not need to be enhanced. Although we are unable at this point to provide statements of sufficient generality, we can trap radiation in a cavity with a Q-ball inside and study numerically its evolution. Our results are summarized in Fig.~\ref{fig:cavity} for a moderately large cavity. Interestingly, they indicate that the system relaxes to a state with a Q-ball that overall has \emph{absorbed} energy, even when starting with a state that initially extracts energy from the soliton (c.f. lower panel of Fig.~\ref{fig:cavity}).
We have not investigated trapped radiation around fundamental solitons in full generality, neither have we studied them when orbited by a point-like charge, a good proxy for some astrophysical systems. Is it possible for a soliton to loose energy to an orbiting point-like charge, making it out-spiral (something known as floating orbit)? This and other aspects remain open.

\noindent {\bf \em Acknowledgments.}
V.C.\ is a Villum Investigator and a DNRF Chair, supported by VILLUM Foundation (grant no. VIL37766) and the DNRF Chair program (grant no. DNRF162) by the Danish National Research Foundation. V.C. acknowledges financial support provided under the European Union’s H2020 ERC Advanced Grant “Black holes: gravitational engines of discovery” grant agreement
no. Gravitas–101052587. Views and opinions expressed are however those of the author only and do not necessarily reflect those of the European Union or the European Research Council. Neither the European Union nor the granting authority can be held responsible for them.
R.V. is supported by grant no. FJC2021-046551-I funded by MCIN/AEI/10.13039/501100011033 and by the European Union NextGenerationEU/PRTR. R.V. also acknowledges support by grant no. CERN/FIS-PAR/0023/2019.
Z.Z.\ acknowledges financial support from China Scholarship Council (No.~202106040037).
This project has received funding from the European Union's Horizon 2020 research and innovation programme under the Marie Sklodowska-Curie grant agreement No 101007855.
We acknowledge financial support provided by FCT/Portugal through grants 
2022.01324.PTDC, PTDC/FIS-AST/7002/2020, UIDB/00099/2020 and UIDB/04459/2020.

\bibliography{ref}

\end{document}